\input epsf
\documentstyle[aps,prb]{revtex}

\def\tens#1{\underline{\underline{#1}}}
\begin{document}

\title{Photoelasticity of sodium silicate glass from first principles}
\author{D. Donadio$^*$, M. Bernasconi}
\address{Dipartimento di Scienza dei Materiali and Istituto Nazionale per la  Fisica
della Materia,  Universit\`a di  Milano-Bicocca, Via Cozzi 53, I-20125, Milano, Italy}
\author{F. Tassone}
\address{Pirelli Cavi e Sistemi Telecom S.p.a., Viale Sarca 222, I-20126, Milano, Italy}
\maketitle
\begin{abstract}
Based on density-functional perturbation theory we have computed 
the photoelastic tensor of a model of sodium silicate glass of composition  
(Na$_2$O)$_{0.25}$(SiO$_2$)$_{0.75}$ (NS3). 
The model (containig 84 atoms) is  obtained by quenching from the melt in combined
classical and Car-Parrinello molecular dynamics simulations.
The calculated photoelastic coefficients are in good agreement with experimental data.
In particular, the calculation reproduces quantitatively the decrease of the
photoelastic response induced by the insertion of Na, as measured experimentally.
 The extension to NS3 of a phenomenological model developed in a previous work                                  
for pure a-SiO$_2$ indicates that the modulation upon strain of  other structural parameters
besides the SiOSi angles must be invoked to explain the change in the photoelstic response induced by Na.
\end{abstract}

\section{Introduction}

Photoelasticity is of interest from the fundamental point of view as well as for several 
technological applications in optics and microelectronics.
For instance, photoelasticity in a-SiO$_2$ is known to cause a reduction of                                              
fiber Bragg gratings efficiency \cite{limberger96} and to produce a loss of resolution                                   
of pure silica lenses used in photolytography \cite{schenker97}. Moreover, photoelasticity 
is directly related to the Rayleigh scattering coefficient in single-component glasses
through the Landau-Placzek relation, and is thus one of the main contributions to loss in 
state-of-the-art silica fibers for telecommunications \cite{schroeder77}. 
A systematic experimental study of photoelasticity in pure and modified silica
glass has been reported by Schroeder in the early 80s.\cite{schroder80}
Brillouin scattering measurements have shown that the modification of silica glass with
alkali or alkali-earth ions (Li, Na, K, Ca, and Mg) reduces sizably the photoelastic 
coefficients.\cite{schroder80}
This result is in contrast with  the prediction of simple models (Lorenz-Lorentz) based on the observed increase
in density and refractive index with the modifier concentration. As expected, a related reduction of the Rayleigh 
scattering coefficient was later measured for these glasses,\cite{schroeder77} and Lines thus 
suggested their use as ultra-low loss glasses for telecommunication applications.\cite{lines88}   

In a recent work \cite{nostro} we have 
shown that the photoelastic coefficients of crystalline and amorphous pure SiO$_2$ can be computed
with good accuracy  within  density functional perturbation theory. A phenomenological model
based on ab-initio data allowed us to identify the microscopic parameters which rule photoelasticity in pure a-SiO$_2$.
In the present paper we have applied the same ab-initio framework to compute the photoelastic tensor of a  
sodium silicate glass with composition (Na$_2$O)$_{0.25}$(SiO$_2$)$_{0.75}$ (NS3) aiming at identifying  
which modification  either structural or electronic induced by the insertion of sodium is mostly responsible 
for the change in photoelastic response.
Models of  NS3 glass  containing  84 atoms have been
generated by quenching from the melt in combined classical and Car-Parrinello molecular dynamics simulations.
The calculated photoelastic coefficients are in good agreement with experimental data. 
In particular, the calculation reproduces quantitatively the changes of the
photoelastic response induced by the insertion of Na, as measured experimentally.
 In order to identify the microscopic mechanims  responsible for the change of the photoelastic response induced by Na,
we  extended to NS3   the phenomenological model developed for pure a-SiO$_2$ in ref. \cite{nostro}.
The paper is organized as follows. In section II we describe our computational framework. 
 In section III we  report the details of  the structural and elastic properties of our model of NS3.         
 The results on the calculated  photoelastic tensor and their interpretation in terms of a
 phenomelological model are presented in sections IV and V, respectively. Section VI is devoted to discussion and conclusions.

\section{Computational details}

A model of sodium silicate glass of composition (Na$_2$O)$_{0.25}$(SiO$_2$)$_{0.75}$
(NS3) has been generated within 
a combined classical and ab-initio framework which has been previously used to generate models
of pure a-SiO$_2$ \cite{nostro,benoit00}. 
Models of the glass have been generated by quenching from the melt in classical
molecular dynamics simulations and then annealed for few ps within ab-initio Car-Parrinello
molecular dynamics \cite{CPMD35}.
The same  method has been used  by Ispas {\sl et al.} \cite{ispas01} to generate  a theoretical model
of  sodium tetrasilicate glass ((Na$_2$O)$_{0.2}$(SiO$_2$)$_{0.8}$, NS4) achieving good agreement with 
experiments both in the structural and electronic properties. 
Ispas {\sl et al.} \cite{ispas01} used a modified BKS-potential 
\cite{vanbeest} extended to sodium silicates by Horbach {\sl et al.} 
\cite{horbach99}. Conversely, 
we have adopted the empirical 
potential developed by Oviedo {\sl et al.} \cite{oviedo98}
which is an extension to sodium silicate glass of the  forcefield introduced by Vashista {\sl et al.} \cite{vashishta}
for pure amorphous silica. It  consists of  
a short-range two-body interaction,  long-range coulomb interactions, and a 
three-body term which enforces the directionality of the  Si-O covalent bond.
The Na-Si, Na-O and Na-Na interactions have been modeled by Oviedo {\sl et al.}
\cite{oviedo98} by coulomb interactions plus a 
repulsive short-range term. The effective charges of the coulombic potential are:
1.0 for Na, 1.6
for Si, and  -0.971 for oxygen which enforces charge   neutrality
in NS3 (note that there is a misprint in the oxygen charges in  table I
of ref. \cite{oviedo98}).
A model of liquid NS3 has been prepared by inserting 7 Na$_2$O  and 21 SiO$_2$ units in 
a cubic box ($a$=10.531 \AA) at the experimental density of glassy NS3 (2.427 g/cm$^3$).
The system is equilibrated at 6800 K, 
cooled to 3800 K in 50 ps and 
then equilibrated at the final temperature for 5 ns. The sample is then quenched
at room temperature in  5 ns  (quenching rate of 7$\cdot$10$^{11}$ K/s) and equilibrated 
at 300 K for 50 ps. This model has then been
annealed at 600 K in Car-Parrinello simulations for 1.3 ps and then quenched  at 300 K in 0.15 ps.
Structural properties have been averaged over a NVE run at room temperature, 1.1 ps long. 
The ab-initio simulations are based on 
density functional theory in the local density approximation (LDA) \cite{pz81} as 
implemented in the code CPMD \cite{CPMD35}.
Norm-conserving 
pseudopotential for Si and Na have been used. Non linear core 
corrections are included in Na pseudopotential \cite{nlcc}. 
An ultrasoft pseudopotential \cite{vanderbilt} has been used for oxygen. 
Kohn-Sham (KS) orbitals are expanded in plane-waves 
up to a kinetic cutoff of 27 Ry.
Integration of the BZ has been restricted to the
$\Gamma$ point. To study the dielectric properties,
the structures generated by Car-Parrinello simulations have been then optimized with
norm-conserving  pseudopotentials and a larger cutoff of 70 Ry.
We have computed the dielectric  and photoelastic tensors within density functional
perturbation theory  DFPT \cite{baroni01}, 
as implemented in
 the code PWSCF and PHONONS \cite{Pwscf}.
The photoelastic tensor is defined by

\begin{equation}
 \Delta\varepsilon^{-1}_{ij}=p_{ijkl}\eta_{kl}
 \end{equation}

 where $\varepsilon_{ij}$ are the components of the optical
 dielectric tensor, $\eta_{kl}$ is the strain tensor.
 Only the electronic contribution is included in $\varepsilon_{ij}$, also indicated as
 $\varepsilon^{\infty}$. Experimentally, it corresponds to the dielectric response
 measured for frequencies of the applied field much higher than
 lattice vibrational frequencies, but lower than the
 frequencies of the electronic transitions.
 The photoelastic coefficients have been calculated by finite differences
 from the  dielectric tensor
 of  systems with strains from -1 $\%$
 to +1 $\%$. 
 The components of the photoelastic
 tensor will be expressed hereafter in  the compressed Voigt notation.

The  exchange-correlation functionals available in literature (LDA and generalized gradient
approximation (GGA)) usually underestimate the
electronic band gap and  overestimate the electronic dielectric tensor up to
10-15 $\%$ \cite{detraux01}.
This discrepancy can be corrected semi-empirically by applying a self-energy
correction, also referred to as a scissor correction, which consists of  a rigid shift of
the conduction bands with respect to the valence bands \cite{scissor}.
This procedure has been used successfully to reproduce the photoelasticity of
Si \cite{levine92}, GaAs \cite{raynolds95} and quartz \cite{detraux01}.
However, as shown in our previous work \cite{nostro}, it turns out that even within simple LDA, 
the error in the photoelastic
coefficients for several polymorphs of silica is  smaller than what expected on the basis of
the error in the dielectric constant itself.
The scissor correction has thus been neglected in the calculation on NS3 reported here.

\section{Structural properties}

\begin{figure}
\centerline{\epsfxsize= 4.3 truecm\epsffile{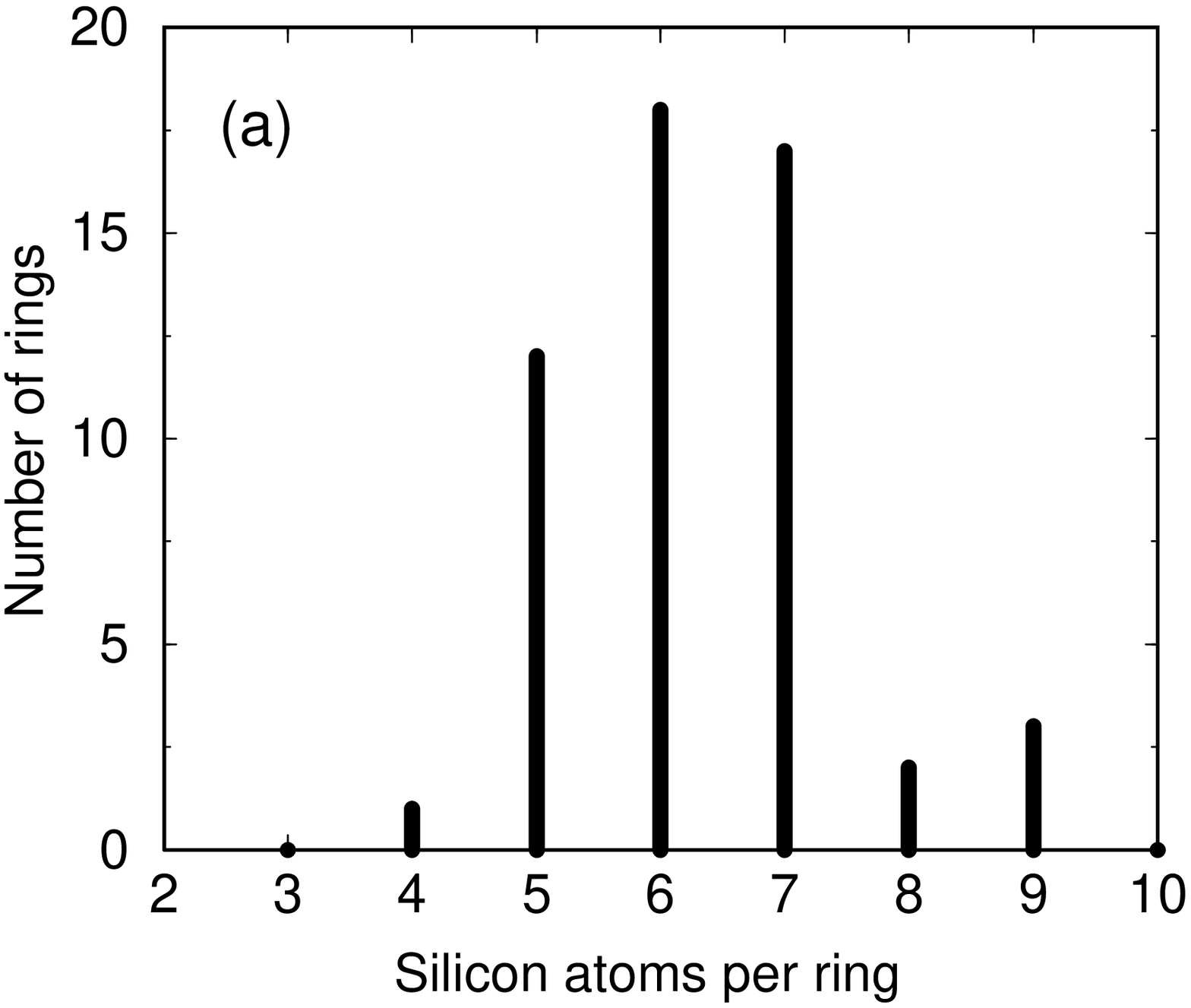}}
\centerline{\epsfxsize= 4.3 truecm\epsffile{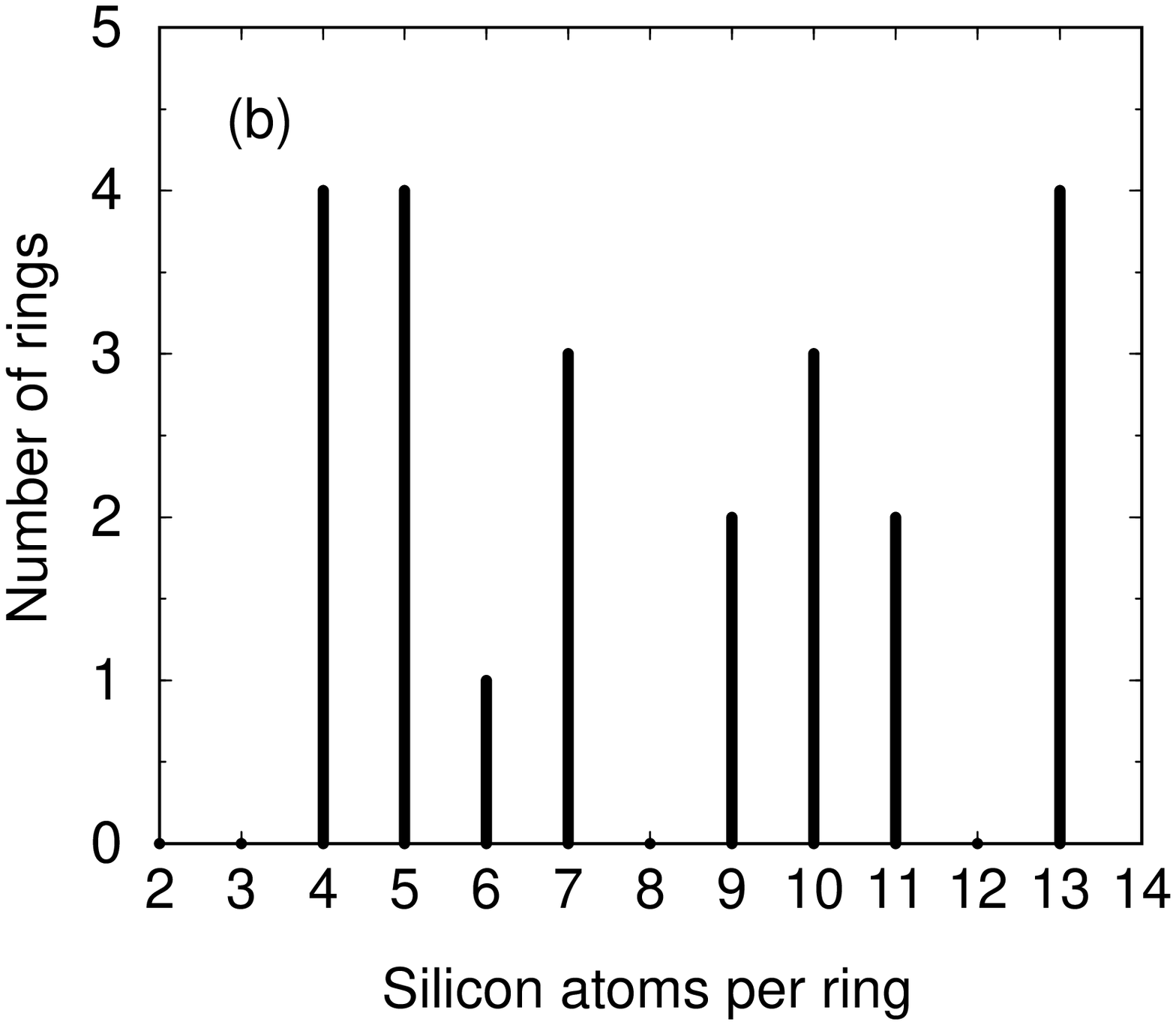}}
\caption{ The ring size distribution of (a) a model of pure a-SiO$_2$ \protect\cite{nostro} and of
(b) the NS3 model after  CPMD annealing.}
\label{ringns}
\end{figure}

As pointed out in previous works \cite{oviedo98,ispas01,horbach99}, 
the insertion of
sodium  modifies the topology of the  network by the 
formation of non-bridging oxygens (NBO). The number of NBO  coincides roughly (and in our small cell, exactly)
with the number of Na atoms.
All silicon atoms in our model are  4-fold coordinated and bonded at most with one NBO.
The presence of NBO drastically
reduces the number of medium sized rings (5-8 Si atoms per ring) 
which are usually predominant in the ring-size distribution of pure silica (Fig. \ref{ringns}).   

\begin{figure}
\epsfxsize= 8. truecm
\centerline{\epsffile{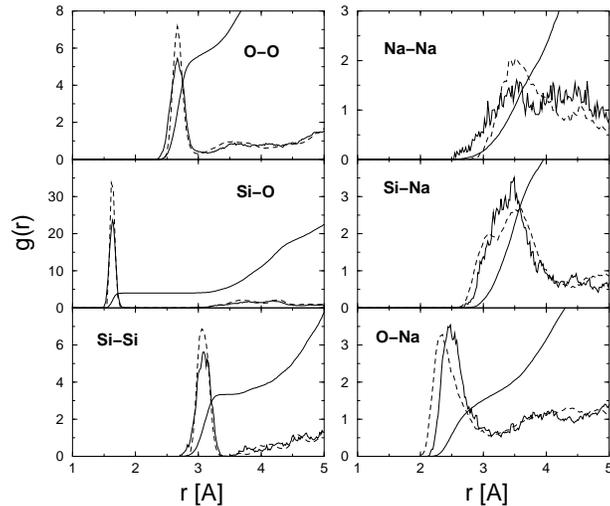}}
\caption{Pair correlation functions of the NS3 model. Bold solid and dashed lines 
         correspond to
         ab-initio and classical MD simulations, respectively. The thin solid 
         line is the ab-initio radial coordination number.}
\label{corrns}
\end{figure}

\begin{figure}
\epsfxsize= 8. truecm
\centerline{\epsffile{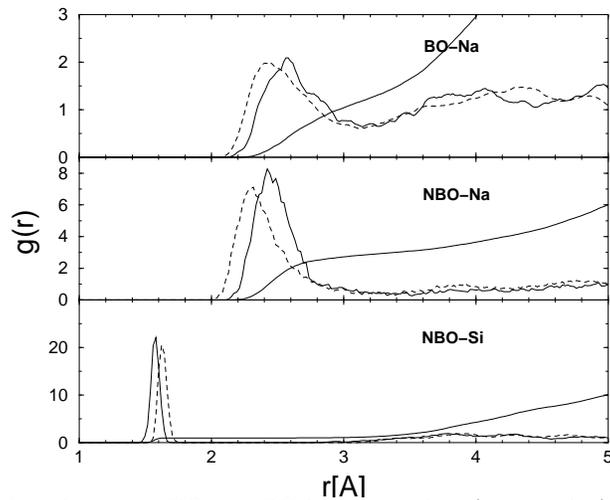}}
\caption{Partial pair correlation functions for NBO and BO in the classical (dashed line) and ab-initio
(bold line) models of NS3.}
\label{corrnbo}
\end{figure}

\begin{figure}
\epsfxsize= 8. truecm
\centerline{\epsffile{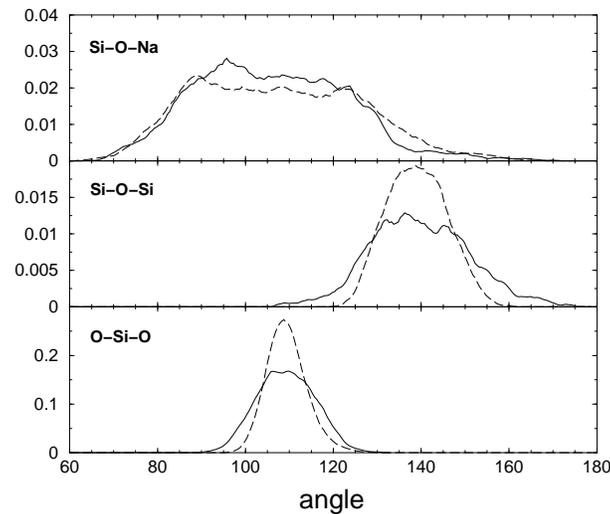}}
\caption{Angular distribution functions. Bold solid and dashed lines
         correspond to
         ab-initio and classical MD simulations, respectively.}
\label{angNS3}
\end{figure}

\begin{figure}
\epsfxsize= 7. truecm
\centerline{\epsffile{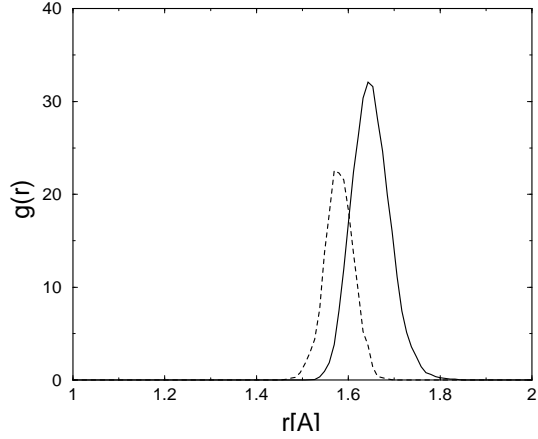}}
\caption{Silicon-oxygen pair correlation function for NBO (dashed line) and BO (solid line) of the ab-initio model of NS3.}
\label{nbo_bo_si}
\end{figure}

The structural properties of the final ab-initio model of  NS3 are compared to 
those of the classical model (before ab-initio annealing) in  
Figs.  \ref{corrns}, \ref{corrnbo} and
\ref{angNS3}.
Pair correlation functions (g(r)) and radial coordination numbers are shown in 
Fig.~\ref{corrns}. 
The average Si-O bond for NBO is slightly shorter than for bridging oxygen ions (BO) as shown in Fig.~\ref{nbo_bo_si}
which report  $g_{SiO}(r)$ resolved for BO and NBO.

The ab-initio annealing produces a slight broadening of the pair correlation
and angle distribution functions (ADF,  Fig. \ref{angNS3}). In fact,  the three-body term in the
classical potential  assigns a  tetrahedral geometry too stiff with respect to the
ab-initio results.
The O-Si-O ADF is still centered around the tetrahedral angle (109.5$^o$), 
while the Si-O-Si ADF is centered at
139.3$^o$, which is smaller than the average Si-O-Si angle in pure
silica.
Furthermore, in contrast with the classical models of pure silica
generated with the BKS potential (cfr. ref. \cite{benoit00,ispas01,nostro}) no shift in the maximum of the $g(r)$ (SiO, OO and SiSi)
is observed upon ab-initio annealing. In fact, the Vashista potential  is fitted directly
on the experimental $g(r)$ of amorphous silica  and the agreement with ab-initio
bondlengths is better than for the BKS potential.
However,  the ab-initio annealing
shifts outwards the peak of the Na-O correlation function  (from 2.4 to 2.5 \AA).
By separating the contribution of 
NBO and BO to the pair correlation functions 
(Fig.~\ref{corrnbo}) we can see that the Na atoms are closer to NBO 
 than to BO. On average, a NBO  is coordinated  with 2.9 Na ions  and a BO with 1.3 Na ions.
On the other hand, a Na ion is on average coordinated with about 2.9 NBO and 3.35 BO
(cfr. Fig.~\ref{corrns}).

Looking at the Na-Na pair correlation function it is worth noting  that 
the broad peak at 3.5 \AA\ present in the classical MD model, disappears
in the CPMD simulation. Thus,
sodium does not tend to segregate or to form clusters, but it is 
homogeneously distributed in the network.

Our results on the structural properties of NS3 are close to those obtained by Ispas et al \cite{ispas01}
for a model of NS4 within a similar theoretical framework.

After the ab-initio annealing  performed with the softer Vanderbilt pseudopotentials,
the final structure
has been further optimized  with norm-conserving pseudopotentials (for the
calculations of the dielectric properties).
 The cell geometry has been optimized
 at fixed volume  allowing orthorhombic distortions
 of the initially cubic supercell such as to produce a diagonal  stress tensor.
 The residual anisotropy in the stress ($\sigma$) 
for the  optimized ratios of cell edges  $b/a=1.04$ and $c/a=1$ is
\begin{equation}
\sigma=\left(\begin{array}{ccc}
          -502.8  &   -10.9 &    4.03 \\
           -10.9  &  -495.1 &   -18.6  \\
            4.03  &   -18.6 & -503.1 \\
         \end{array}  \right)\ \ \ kbar
\label{sigma}
\end{equation}

\begin{figure}
\epsfxsize= 8. truecm
\centerline{\epsffile{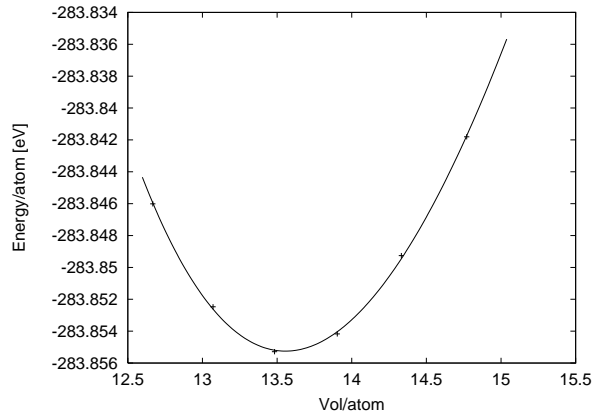}}
\caption{Ab-initio equation of state of the NS3 model.} 
\label{eosNS3}
\end{figure}

The large negative stress in eq. \ref{sigma} is  due to the so--called Pulay stress.
The $b/a$ and $c/a$ ratios  obtained in this way at the experimental equilibrium density 
have been  held fixed
and the volume varied to generate the equation of state reported in Fig.
\ref{eosNS3}.

The calculated $E(V)$ points have been corrected for the discontinuities
due to the incomplete basis set following the prescription given in ref. \cite{francis90} and
then fitted by a Murnaghan function \cite{murna}. The resulting equilibrium density ($\rho_{eq}$),
 bulk modulus ($B$) and derivative of the bulk modulus with respect to pressure ($B'$)
are $\rho_{eq}=2.468$ g/cm$^3$ (exp. $2.427$ g/cm$^3$), $B=44.1$ GPa,
$B'=1.6$. 
The insertion of Na produces a  marginal softening of $B$ with respect to pure silica \cite{nostro}.
In fact in our previous work \cite{nostro} on a model of a-SiO$_2$ of similar size (81 atoms) we obtained in our 
previous work \cite{nostro}
we have obtained $\rho_{eq}=2.31$ g/cm$^3$, $B=46.2$ GPa,
$B'=-6.28$. 
 The value of $B'$, negative in pure silica,
turns to positive in NS3, which means that the structural response to
compression is different in silica and in NS3. 
Na-O interaction is probably responsible for the change in $B'$.

 The response to strain 
($\eta_{22}$,$\eta_{33}$) of some average structural 
parameters of NS3 and of a-SiO$_2$ (81-atoms supercell) are compared in table \ref{str_res}.
The change in the SiOSi angles upon strain is substantially smaller in NS3 than in pure a-SiO$_2$, 
nevertheless the response of the Na-O distances to strain is rather relevant, if compared to the average 
Si-O distance changes, both in pure silica and in NS3.

\begin{table}
\caption{The derivatives of the SiO and NaO bond-lengths, SiOSi angles and  Si-Si nearest neighbors vector distances
         with respect to strains $\eta_{22}$ and $\eta_{33}$ in the models of NS3 and a-SiO$_2$ \protect\cite{nostro}.
         P$_i$(Si-Si) denotes the projection of the average Si-Si vector distance on the $i$-th axis. NBO and BO indicate non-bridging
	 and bridging oxygen atoms, respectively.}
\label{str_res}
\begin{center}
\begin{tabular}{lcc|cc}
 &         NS3    &               &  SiO$_2$         &   \\ 
 &  $\partial /\partial\eta_{22}$ & $\partial /\partial\eta_{33}$  & $\partial /\partial\eta_{22}$  & $\partial /\partial\eta_{33}$ \\
\hline
  $\overline{\rm Si-O}$    &  0.145    & 0.125   & 0.154 & 0.157    \\
  $\overline{\rm{Si-}\widehat{\rm O}{\rm -Si}}$ & 49.7    & 39.5   & 69.1 & 73.4  \\
  P$_x$(Si-Si) & -0.218  & -0.290  & -0.037   & -0.122 \\
  P$_y$(Si-Si) &  1.451  & -0.131  &  1.553   & -0.049 \\
  P$_z$(Si-Si) & -0.026  &  1.303  & -0.067   &  1.490 \\
  $\overline{\rm Na-BO}$ &  0.790  &  0.676   & - & - \\
  $\overline{\rm Na-NBO}$\ \ \ \ & 0.479   &  0.280   & - & - \\
\end{tabular}
\end{center}
\end{table}

\section{Dielectric and photoelastic properties}

The dielectric tensor of the NS3 model optimized at its equilibrium
density is:
\begin{equation}
\varepsilon=\left(\begin{array}{ccc}
          2.449   &  0.008  & -0.006 \\
          0.008   &  2.438  &  0.031 \\
         -0.006   &  0.031  &  2.474 \\
         \end{array}  \right)\ \ \ ,
\label{epsiNS}
\end{equation}
The average theoretical dielectric constant of NS3 and of pure silica (cfr. Ref.$[6]$)
are compared to experimental data in table .
\begin{table}
\caption{Theoretical and experimental dielectric constant of NS3 and pure a-SiO$_2$. $^a$ Ref.$[6]$ .}
\label{epsiNS3}
\begin{center}
\begin{tabular}{lcc}
&  DFPT  &   exp.\protect\cite{schroder80}      \\
\hline
   NS3           & 2.454 &  2.236   \\
   a-SiO$_2$ \ \ \ \    & 2.292$^a$ &  2.125   \\
\end{tabular}
\end{center}
\end{table}
The increase of $\varepsilon$ due to  sodium insertion is quantitatively reproduced by our
calculations.
The computation of the photoelastic tensor has been performed 
by finite differences of the dielectric tensor by applying
strains of $\pm 1 \%$. 

Results on the photoelastic coefficients are compared with experiments \cite{schroder80}  in table \ref{photoNS3}.
The $p_{11}$ and $p_{12}$ coefficients have been obtained by averaging over different components
which should be equal in a homogeneous model, i.e. $p_{11}=(p_{33}+p_{22})/2$ and 
$p_{12}=(p_{12}+p_{13}+p_{23}+p_{32})/4$.
The agreement with experimental data is of the same quality as for pure a-SiO$_2$.
In particular, the calculation reproduces quantitatively the change in
photoelastic coefficients observed experimentally upon insertion of Na,  namely
a large decrease of the off-diagonal $p_{12}$ and a smaller increase of $p_{11}$. 
\begin{table}
\caption{Ab-initio photoelastic coefficients of the NS3 model
compared with experimental data, with photoelastic coefficients of pure silica computed by DFPT and
with measured ones.}
\label{photoNS3}
\begin{center}
\begin{tabular}{lcc|cc}
           & \ \ \ \ \ NS3 &   &\ \ \ \ \ \ \ a-SiO$_2$   &   \\             
           & This work &  exp. \cite{schroder80} & DFPT \cite{nostro} & exp.  \cite{schroder80} \\
\hline
$p_{11}$   &  0.097  &  0.134      & 0.057  & 0.125  \\
$p_{12}$   &  0.167  &  0.214      & 0.220  & 0.27   \\
$p_{44}$   & -0.044  &  -0.040     & -0.074 & -0.073 \\
\end{tabular}
\end{center}
\end{table}

\section{Phenomenological model of photoelasticity}

\subsection{Pure silica glass}

In a previous work \cite{nostro}, we have developed a phenomenological model
of the dielectric properties of silica based on ab-initio data.
The main features of the model are briefly outlined here. Its  extension to NS3 is described in the next section.
In ref. \cite{nostro} we have assumed that 
the dielectric response of silica polymorphs could be embodied in an ionic polarizability
tensor of the oxygen ions, whose value is assumed to depend on the Si\^OSi angle only.
The dielectric susceptibility  $\tens{\chi}$ ($\tens{\epsilon}=1+ 4 \pi \tens{\chi}$)
can be obtained  from a site dependent
oxygen polarizability, $\tens{\alpha_i}$, as 

\begin{equation}
  \tens{\chi} = \frac{1}{V}\sum_{i,j}^N \tens{\alpha_i} \left( \tens{I}-\tens{B} \right)_{ij}^{-1},
  \label{chi}
  \end{equation} 

where $V$ is the volume of the unit cell containing $N$ oxygen ions and
$\tens{B}$ is a 3Nx3N matrix consisting of 3x3 blocks $\tens{B_{ik}}$ defined as
\begin{equation}
  \tens{B_{ik}}=\left(\frac{4\pi}{V}-\sum_{\vec R}\tens{T^{\vec R}_{ik}}\right)\tens{\alpha_k}\ \ \,
  \end{equation}

and 

\begin{equation}
  \tens{T^{\vec{R}}_{ik}} =\nabla_i\nabla_k\left(\frac{1}{r_{ik}}\right)=
    \frac{1}{r_{ik}^3}\left[1-3\frac{\vec{r}_{ik}\vec{r}_{ik}}{r_{ik}^2}\right],
    \end{equation}
    where $\vec{R}$ are Bravais lattice vectors 
    defined by the shape of the supercell for models of  glasses or by the unit cell for crystalline phases.
    $\vec{r}_{ik}$
    is the distance between sites $i$ and $k$ in cells separated by  $\vec{R}$ (see Ref. \cite{nostro} for details).

\begin{figure}[!ht]
\centerline{\epsfxsize= 7. truecm
\epsffile{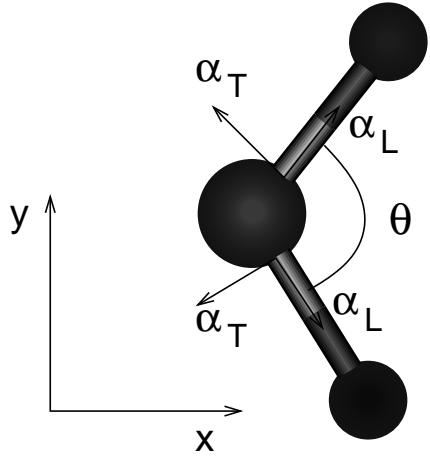}}
\caption{Sketch of  the Si-O-Si unit and of the bond contributions to polarizability.}
\label{modelfig}
\end{figure}

\begin{figure}[!ht]
\centerline{\epsfxsize= 7. truecm \epsffile{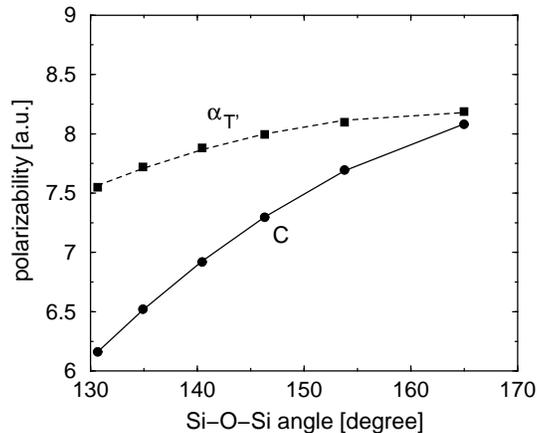}}
\caption{The functions $c(\theta)$ (solid line and circles) and $\alpha_{T'}(\theta)$ 
(dashed line and squares), which assign the oxygen polarizability (see text). The data are the result
 of the fitting  on the dielectric tensor of $\alpha$-cristobalite
 at different densities.} 
\label{alphac}
\end{figure}

The polarizability of the oxygen ions  is a function of the SiOSi angle $\theta$ and can be expressed in terms
of the polarizability of the SiO bonds as (cfr. ref. \cite{nostro})
\begin{equation}
 \tens{\alpha}= \left( \begin{array}{ccc}
    c(\theta)+\gamma cos^2(\theta /2) &   0  &  0 \\
       0  &    c(\theta)+\gamma sin^2(\theta /2) &  0 \\
	  0 & 0 & \alpha_{T'}(\theta)
	     \end{array} \right)
	     \label{polar}
	     \end{equation}

	     with $c=2(\alpha_L + \alpha_T + \alpha_{T'})/3$, $\gamma=\alpha_L -\alpha_T$
and $\alpha_L$,  $\alpha_T$ and $\alpha_{T'}$ are  the  longitudinal and the two transversal polarizabilities of
the Si-O bond in Fig. \ref{modelfig}, respectively.
    The contribution of each polarizable SiOSi unit to the dielectric susceptibility is 
	     \begin{equation}
	      \tens{\alpha_i} = \tens{R_i^T}\tens{\alpha}(\theta)\tens{R_i}
	      \end{equation}
	      where $\tens{R_i}$ is the rotation matrix that operates the transformation from 
	      the local reference system represented in figure
	      \ref{modelfig} to the absolute reference system of the solid, in which the $i$-th Si-O-Si unit is embedded.
The parameter 
$\gamma$ has been assumed independent on $\theta$ and
 set equal to the value  obtained from the fitting of the  ab-initio Raman spectrum of $\alpha$-quartz
 in ref. \cite{umari01} ($\gamma = 9.86$ a.u.).
As discussed in our previous work \cite{nostro}, the functions $c(\theta)$ and $\alpha_{T'}(\theta)$ have been fitted on
the dielectric properties of  $\alpha$-cristobalite at different densities.  The results  reported are shown in Fig. \ref{alphac}.

This model allows the calculation of the photoelastic tensor given  by the derivative of the 
dielectric susceptibility tensor $\chi$ (eq. \ref{chi}) with respect to the strain tensor 
$\eta_{\lambda\mu}$  as
\begin{eqnarray}
  \frac{\partial\tens{\chi}}{\partial\eta_{\lambda\mu}} = -\tens{\chi}\delta_{\lambda\mu} +  
	\frac{1}{V} \sum_i^N \Bigg(\left[ \tens{R_i^T}\frac{\partial\tens{\alpha}}{\partial\eta_{\lambda\mu}}\tens{R_i}\right]
	      \tens{A} + \nonumber\\
		   +\left[\tens{R_i^T}\tens{\alpha}\tens{R_i}\right]\frac{\partial\tens{A}}{\partial\eta_{\lambda\mu}}+
		    \left[\frac{\partial \tens{R_i^T}}{\partial\eta_{\lambda\mu}} \tens{\alpha} \tens{R_i} 
		     + \tens{R_i^T}\tens{\alpha}\frac{\partial \tens{R_i}}{\partial\eta_{\lambda\mu}}\right]\tens{A}\Bigg)
		     \label{dchi}
		     \end{eqnarray}
		     where the matrix $\tens{A}$ is defined as $\tens{A}=(\tens{I}-\tens{B})^{-1}$ (see eq. \ref{chi}), 
		     and the arguments in square brackets indicate
		     3x3N matrices.
		     The change of the oxygen polarizability with strain can be expressed in terms of
		     $\partial c(\theta)/\partial\theta$ and $\partial\alpha_{T'}(\theta)/\partial\theta$ deduced from Fig. \ref{alphac}.
		     All the other derivatives can be  obtained by finite differences.
The phenomenological model  outlined  above has been shown  to reproduce 
satisfactorly the dielectric and photoelastic tensors of several silica polyphorms
(quartz, $\alpha$-cristobalite,
$\beta$-cristobalite, a-SiO$_2$) at normal conditions and at high density \cite{nostro}.

\subsection{Extension to sodium silicates}

We can now make use of this phenomenological model for pure a-SiO$_2$
to identify which term in Eq. \ref{dchi} is mostly affected by the presence of sodium.
As shown in Ref. \cite{schroder80} the change in $\varepsilon$ and in the density alone 
within a simple Lorenz-Lorentz model (cfr. section II) can not account for
the change in the photoelastic coefficients measured experimentally upon Na insertion.
This is inferred by comparing the quantity ${\rho\partial\varepsilon/\partial\rho}$
measured experimentally ($(\rho\partial\varepsilon/\partial\rho)_{obs}=\varepsilon^2(p_{11}+2p_{12})/3$)
with the result of the Lorenz-Lorentz model 

\begin{equation}
(\rho\frac{\partial\varepsilon}{\partial\rho})_{LL}=\frac{(\varepsilon-1)(\varepsilon+2)}{3}
\label{LL}
\end{equation}

For pure a-SiO$_2$, $(\rho\partial\varepsilon/\partial\rho)_{obs}$=1.003 and
$(\rho\partial\varepsilon/\partial\rho)_{LL}$=1.547, while for NS3 
$(\rho\partial\varepsilon/\partial\rho)_{obs}$=0.937 and  $(\rho\partial\varepsilon/\partial\rho)_{LL}$=1.746.
Thus, the Lorenz-Lorentz model predicts an increase in the photoelastic response with Na content
which is in contrast with the experimental observation.
An increase in the dependence of the molecular polarizability upon strain 
($\partial\alpha/\partial\eta$ in Eq. \ref{dchi}),
neglected in the Lorentz-Lorenz model,  is required to account for a decrease in $p_{12}$.
 By still keeping valid the phenomenological
 model of pure silica, we may argue that the quantity $\partial\alpha/\partial\eta$
 may change with Na content  because of different possible effects: i) a change of
the elastic response of the system (via $\partial\theta/\partial\eta$), ii) a change in the
functional dependence of the BO polarizability on the SiOSi angle  (cfr. Fig. \ref{alphac}),
 iii) other structural parameters which become relevant with the presence of Na (such as the NaO interaction for instance)
control the response to strain of the polarizability of BO, iv) the NBO contribute to a term in $\partial\alpha/\partial\eta$.
In ref. \cite{lines88} Lines has conjectured that the decrease in the photoelastic coefficients  in NS glasses might be due to an increased
modulation of the SiO bond length by strain, i.e. the Na ions would simply modify the mechanical response of the glass.
However, we can recognize that this conjecture is not supported by the results in table I. In fact, the modulation of the SiO
bond length upon even decreases in NS3 with respect to pure silica (cfr. table I). 
In order to clarify these issues, we have extended the phenomenological model for pure silica  by
considering different polarizabilities for NBO and BO. We have introduced an additional parameter, $\alpha_{NBO}$, which describes
a spherical polarizability of NBO ions. 
We have further assumed that $\alpha_{NBO}$ may depend on the local electric field ($E_{loc}$) produced by the Na$^+$ ions.
The modulation of the Na-NBO distances upon strain  
 would thus contribute to $\partial\alpha/\partial\eta$ in the
photoelastic tensor  via  a term of the form
\begin{equation}
\frac{\partial\alpha_{NBO}}{\partial\eta}=\frac{\partial\alpha_{NBO}}{\partial E_{loc}}\frac{\partial E_{loc}}{\partial\eta}\ .
\label{Eloc}
\end{equation}
We have fitted the parameter ${\partial\alpha_{NBO}}/{\partial E_{loc}}$, which describes the response of the NBO polarizability 
to the local field of Na$^+$
on the photoelastic coefficients of a sodium silicate crystal: the
natrosilite Na$_2$Si$_2$O$_5$ \cite{natro}.

\subsubsection{Natrosilite}

Natrosilite is a  layered material with NBO nearly aligned  along the $c$-axis, perpendicular to the  siloxane layers (Fig. 1).
Details on the structure of natrosilite are given in the appendix.

\begin{figure}
\centerline{\epsfxsize= 8. truecm \epsffile{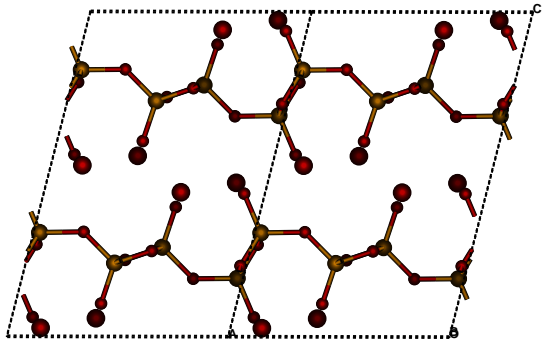}}
\caption{Side view of the natrosilite crystal. Large (small) dark grey spheres are Na (O) ions. Light grey spheres are Si ions.}
\label{natrosilite}
\end{figure}

By changing the length of the  $c$ axis with  the other lattice parameters ($a$, $b$ and $\beta$, see appendix)
fixed, the SiOSi angles of BO do not change while there is
a large change of the Na-NBO distances. We have thus assumed that the change in the dielectric constant of natrosilite 
upon the $\eta_{3}$ strain would be entirely due to the change of $\alpha_{NBO}$ which would obviuosly lead to an
overestimation of $\partial\alpha_{NBO}/{\partial\eta}$.
We have thus considered a model for the dielectric response of natrosilite in which the BO have the same polarizability as in
pure silica (cfr. section Va) and  the two additional parameters of NBO, $\alpha_{NBO}$ and $\partial\alpha_{NBO}/{\partial\eta_3}$,
have been fitted on the ab-initio 
 photoelastic coefficient of natrosilite $p_{13}$, $p_{23}$, and $p_{33}$.
 We obtain $\alpha_{NBO}=14.6\ a.u.$ and $\partial\alpha_{NBO}/{\partial\eta_{3}}= 20.5\ a.u$.
From Eq. \ref{Eloc}

\begin{equation}
\frac{\partial\alpha_{NBO}}{\partial\eta_{3}}=\frac{\partial\alpha_{NBO}}{\partial E_{loc}}\frac{\partial E_{loc}}{\partial\eta_3}=
\frac{\partial\alpha_{NBO}}{\partial E_{loc}} N_{Na} \frac{\partial <r^{-2}_{NaO}>}{\partial \eta_3}
\label{Eloc2}
\end{equation}

where we have assumed 

\begin{equation}
E_{loc}=N_{Na} <r^{-2}_{NaO}>,
\label{Eloc3}
\end{equation}

$N_{Na}$ is the number of $Na$ ions nearest neighbor to NBO and 
$r_{NaO}$ is the Na-NBO distance. The average is over the nearest neighbor ions.
In natrosilite $N_{Na}$=4 and $<r^{-2}_{NaO}>= 0.177$ \AA $^{-2}$ which finally yields 
$\partial\alpha_{NBO}/{\partial E_{loc}}=-29.3\ a.u.$.
Table IV  reports the ab-initio photoelastic  coefficients of natrosilite and the results of the
phenomenological model   outlined above.
The quantitity $\partial\alpha_{NBO}/{\partial E_{loc}}$ is negative which means that by moving the Na ions further away  from the NBO its
polarizability  increases. In fact, by decreasing the local electric field on NBO its charge would become more diffuse and thus more
polarizable.
The ab-initio photoelastic tensor has been computed by finite differences from the dielectric tensor calculated within
DFPT with the codes PWSCF and PHONONS as described in section II. The calculations have been performed at the
experimental equilibrium volume (see appendix).

\begin{table}                                                                                                          
 \caption{Average values of the phenomenological polarizability of BO and NBO ions and                                  
  photoelastic coefficients  of natrosilite computed ab-initio (DFPT) and with the phenomenologial model                 
   described in the text.}                                                                                                
\label{natrosil}                                                                                                       
\begin{center}
\begin{tabular}{lcc}                                                                                                   
  &  Model    &  DFPT \\ 
\hline
 $\overline{\alpha_{BO}}$     & 10.6   &  -     \\                                                                             
 $\overline{\alpha_{NBO}}$\ \ \ \    & 14.62  &  -     \\                                                                          
 $\varepsilon_{11}$           & 2.454  & 2.412  \\
 $\varepsilon_{22}$           & 2.467  & 2.443  \\
 $\varepsilon_{33}$           & 2.282  & 2.361  \\
   $p_{13}$                   & 0.140  & 0.131  \\  
   $p_{23}$                   & 0.143  & 0.138  \\ 
   $p_{33}$                   & 0.087  & 0.104  \\
 \end{tabular}                                                                                                          
\end{center}
\end{table}

\subsubsection{NS3 glass}

\begin{figure}
\centerline{\epsfxsize= 8. truecm \epsffile{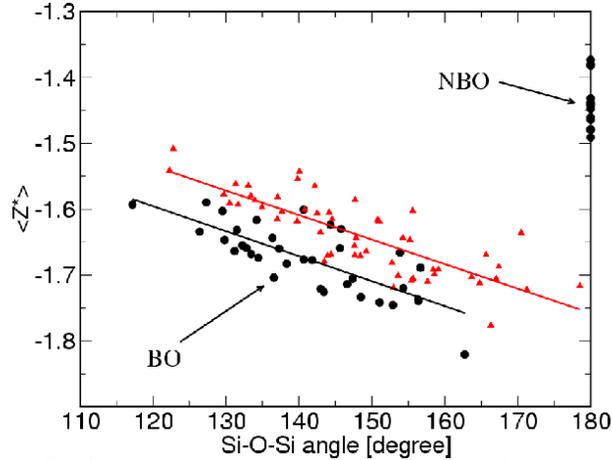}}
\caption{Dependence of the Born effective charges of oxygen atoms as a function of the
         Si-O-Si angle in the models of NS3 (black circles)  and  a-SiO$_2$
         (red triangles, from ref. \protect\cite{nostro}). 
	 BO and NBO indicate
         bridging and non-bridging oxygen ions, respectively.} 
\label{zeffNS3}
\end{figure}

In the development of  a model for photoelasticity in NS3,  we have used the value of 
$\partial\alpha_{NBO}/\partial E_{loc}$ obtained from the fitting on natrosilite described in the previous section.
The differences in the environment of NBO and in the structural response to strain of NS3 with respect to natrosilite
is accounted for by the term 
$\partial E_{loc}/\partial\eta$ in Eq. \ref{Eloc2}. In our model of NS3, $N_{Na}$=2.8 and 
$<r^{-2}_{NaO}>= 0.166$ \AA $^{-2}$ (cfr. section III).
To extend the phenomenological model to NS3, we have considered two extreme cases as described below.

a) In the first case (model A) we have assumed that the polarizability of BO ions ($\alpha_{BO}$) is the same as in pure
a-SiO$_2$ and assigned by the curves in Fig. \ref{alphac}. The polarizability of NBO, $\alpha_{NBO}$, is then assigned by fitting
the ab-initio dielectric constant of NS3 within our extended phenomenologial model. 
The contribution from $\partial\alpha_{NBO}/\partial E_{loc}$, fitted on natrosilite, is added.
The results are reported in table V.
The fitting  yields $\alpha_{NBO}=17.0 \, a.u.$, a value larger than  $\alpha_{NBO}$ in natrosilite (cfr. table IV), as we would have
expected  by considering that $\partial\alpha_{NBO}/\partial E_{loc}$ is negative and $E_{loc}$ is  lower in NS3 with respect to
natrosilite (cfr. Eq. \ref{Eloc3}).
If we change by  20 $\%$ the value of  $\alpha_{NBO}$ (a change comparable with
the difference in $\alpha_{NBO}$ between natrosilite and NS3) in the fitting procedure on natrosilite, one obtains a
value for $\partial\alpha_{NBO}/\partial E_{loc}$ with a similar change of 20 $\%$ with respect to the data in table
\ref{natrosil}. 
However, a change in $\partial\alpha_{NBO}/\partial E_{loc}$ of the order of 20 $\%$ does not affect the
the results on the photoelastic coefficients of NS3 within the figures reported in table V.

b) In the second case (model B) $\alpha_{NBO}$ is set to zero and   $\alpha_{BO}$ is still assigned by
the function given in Fig. \ref{alphac}, but rescaled by a multiplicative factor in order to reproduce the dielectric
constant of NS3. In this way we have also rescaled the  term $\partial\alpha_{BO}/{\partial \eta}$ depending on
the SiOSi angle. 
 The increase of the polarizability of the BO upon Na insertion is supported by the comparison of the 
calculated Born effective
charges for  our models of NS3 and  a-SiO$_2$  shown in Fig. \ref{zeffNS3}.
The presence of Na induces a nearly uniform increase of                     
the Born effective charges of  BO ions. The  valence electrons of the ionized Na atoms     
are thus transferred to both NBO and BO ions. 
A larger charge on the BO ions  implies a larger polarizability.
In this model the contribution from NBO is added as well with
the same parameter $\partial\alpha_{NBO}/\partial E_{loc}$
fitted on natrosilite and used in model A.
The results are reported in table  V.

\begin{table}
\caption{Average values of the phenomenological polarizability of BO and NBO ions and
photoelastic coefficients according to the two models 
(A and B) described in the text. The ab-initio (DFPT) photoelastic coefficients of NS3 and pure a-SiO$_2$
(ref. \cite{nostro}) are also reported. Results including and neglecting the term $\partial\alpha_{NBO}/\partial E_{loc}$
(in a.u., see text) are both reported.}
 \label{alfaNBO}
\begin{center}
\begin{tabular}{lccccc|cc}
 &      & & NS3     &    &           &   \ \ \ \ a-SiO$_2$  &          \\
  &  & Model A & & Model B  & DFPT & Model & DFPT \\
  \hline
 $\overline{\alpha_{BO}}$               &  10.6 & 10.6    & 17.8  & 17.8   &  -  & 10.6  & -    \\
 $\overline{\alpha_{NBO}}$              &  17.0 & 17.0    &    0  & 0      &  - & -  & -  \\
$\partial\alpha_{NBO}/\partial E_{loc}$ &  0    &  -29.33 &    0  & -29.33 &  - & -  & -  \\
                       	$p_{21}$        &  0.25 &  0.24   & 0.24  &  0.23  & 0.167 & 0.227  & 0.220 \\
                	$p_{22}$        &  0.18 &  0.17   & 0.14  &  0.13  & 0.072 & 0.097  & 0.057 \\
\end{tabular}
 \end{center}
\end{table}
Both models A and B largely overestimate the photoelastic coefficients
in NS3 with respect to pure a-SiO$_2$. 

As mentioned before the value used for $\partial\alpha_{NBO}/\partial E_{loc}$  overestimates the real contribution of NBO
to $\partial\alpha/\partial\eta$ in NS3. Yet, although overestimated,  NBO  gives a very small contribution to the
photoelastic tensor.
Even within model B where the contribution of BO is also rescaled by a factor 1.7 by still keeping $\alpha_{BO}$ dependent on the
SiOSi angle only, the  model photoelastic coefficients are sizably  larger than the ab-initio  data.
The failure of these  models suggests that
either the shape of the functions in Fig. \ref{alphac} changes upon Na insertion and/or
other structural parameters, in addition to the SiOSi angle, influence the 
polarizability of BO ions.

\section{Conclusions}

Photoelasticity in a model of  sodium silicate glass (NS3) 
has been studied within  density functional perturbation theory. 
The NS3 model is generated by quenching from the
melt in combined classical and Car-Parrinello molecular dynamics simulations.

The calculated photoelastic coefficients 
are in good agreement with experimental data and further confirm the reliability of DFPT already assessed for pure
silica polymorphs in our previous work \cite{nostro}.
In particular, the calculation reproduces quantitatively the decrease of the
photoelastic response induced by the insertion of Na, as measured experimentally \cite{schroder80}.
Aiming at identifying the microscopic mechanisms  thorough which sodium modifies the photoelastic response of the glass,
we have extended to NS3 a  phenomenological model of photoelasticity  developed for  pure silica
in our previous work \cite{nostro}.
It comes out that the contribution of NBO to the photoelastic tensor (via  
a modulation of the NBO polarizability with strain) is not
large enough to explain the decrease of the photoelastic coefficient of NS3 with respect to that of pure a-SiO$_2$.
 Moreover, although a charge transfer takes place from                                                 
the ionized Na ions to the                                                                                           
BO ions,  a simple increase of  the  polarizability of BO
is  not sufficient to explain the change in the photoelastic response 
 by keeping the SiOSi angle as the only structural parameter
which modules the polarizability upon strain as in pure a-SiO$_2$. The modulation upon strain
of other structural parameters should be called for.
In order to quantify these latter effects, a more detailed modeling of the 
must be devised, 
 requiring the  fitting of the dielectric properties  over a large database of sodium silicate
crystals. 

$^*$ present address: Computational Science, Department of Chemistry and Applied
  Biosciences, ETH Zurich, USI Campus, Via Giuseppe Buffi 13, CH-6900
    Lugano, Switzerland.

\begin{acknowledgments}

D.D. acknowledges Pirelli Cavi e Sistemi S.p.a. for financial support.
\end{acknowledgments}

\appendix

\section{}

Natrosilite, Na$_2$O(SiO$_2$)$_2$, is a phyllosilicate crystal with space group P21/c 
and 4 formula units per unit cell
\cite{natro}.
We have optimized the internal structure at the experimental equilibrium lattice
parameters $a$=8.13 \AA , $b$=4.85 \AA , $c$=12.33 \AA\ and $\beta=104.3^o$\cite{natro}.
We have used the code PWSCF \cite{Pwscf} and  2x2x2 Monkhorst-Pack 
\cite{MP} mesh for Brillouin Zone integration.
The experimental and theoretical positions (in crystallographic  units)  of the 9 independent atoms
are reported in table \ref{posnatro}

\begin{table}[ih]
\caption{Experimental \protect\cite{natro} and theoretical positions (in crystallographic units) 
of the independent atoms of natrosilite.}
\label{posnatro}
\begin{center}
\begin{tabular}{l c c c | c c c}
      &         & Exp.    &          &          &  DFT    & \\
      &    x    &   y     &   z      &     x    &   y     &   z  \\
\hline
  Si &\ \  0.028 \ \ &\ \ 0.184 \ \ &\ \ 0.183 \ \   &\ \  0.027 \ \  &\ \ 0.164 \ \ &\ \   0.181 \ \\ 
  Si &   0.403  &  0.295  &  0.277   & 0.402  &    0.304  &    0.276  \\
  Na &   0.379  &  0.753  &  0.443   & 0.381  &    0.752  &    0.443  \\ 
  Na &   0.137  &  0.225  &  0.473   & 0.143  &    0.210  &    0.475  \\ 
   O &   0.029  &  0.859  &  0.215   & 0.026  &    0.832  &    0.216  \\ 
   O &   0.454  &  0.620  &  0.267   & 0.445  &    0.637  &    0.262  \\ 
   O &   0.226  &  0.246  &  0.181   & 0.225  &    0.242  &    0.179  \\ 
   O &   0.391  &  0.232  &  0.401   & 0.388  &    0.243  &    0.400  \\ 
   O &   0.093  &  0.755  &  0.436   & 0.093  &    0.721  &    0.439  \\ 
\end{tabular}
\end{center}
\end{table}

\end{document}